\documentclass[runningheads]{llncs}
\usepackage{latexsym,amsmath}
\usepackage{amssymb}

\begin{document}


\title{A Note on Parameterised Knowledge Operations in Temporal Logic}


\author{Vladimir Rybakov}
\authorrunning{V.Rybakov}
\titlerunning{A Note on Parameterised Knowledge Operations in Temporal Logic
 }

\institute{School of Computing, Mathematics and DT,
  Manchester Metropolitan University,
 John Dalton Building, Chester Street, Manchester M1 5GD, U.K,
 \email{V.Rybakov@mmu.ac.uk}
 }

\toctitle{Lecture Notes in Computer Science}

\mainmatter

\maketitle






%


\textheight 552.4pt \textwidth 5in


\newcommand{\ga}{\alpha}
\newcommand{\gb}{\beta}
\newcommand{\grg}{\gamma}
\newcommand{\gd}{\delta}
\newcommand{\gl}{\lambda}
\newcommand{\cff}{{\cal F}or}
\newcommand{\ca}{{\cal A}}
\newcommand{\cb}{{\cal B}}
\newcommand{\cc}{{\cal C}}
\newcommand{\cm}{{\cal M}}
\newcommand{\cmm}{{\cal M}}
\newcommand{\cbb}{{\cal B}}
\newcommand{\ccrr}{{\cal R}}
\newcommand{\cf}{{\cal F}}
\newcommand{\cy}{{\cal Y}}
\newcommand{\cxx}{{\cal X}}
\newcommand{\cdd}{{\cal D}}
\newcommand{\cww}{{\cal W}}
\newcommand{\czz}{{\cal Z}}
\newcommand{\cll}{{\cal L}}
\newcommand{\cw}{{\cal W}}
\newcommand{\ckk}{{\cal K}}
\newcommand{\ppp}{{\varphi}}

\newcommand{\ii}[0]
{\rightarrow}

\newcommand{\ri}[0]
{\mbox{$\Rightarrow$}}

\newcommand{\lri}[0]
{\mbox{$\Leftrightarrow$}}

\newcommand{\lr}[0]
{\mbox{$\Longleftrightarrow$}}

\newcommand{\ci}[1]{\cite{#1}}

\newcommand{\pr}{{\sl Proof}}

\newcommand{\vv}[0]{
\unitlength=1mm
\linethickness{0.5pt}
\protect{
\begin{picture}(4.40,4.00)
\put(1.2,-0.4){\line(0,1){3.1}}
\put(2.1,-0.4){\line(0,1){3.1}}
\put(2.1,1.1){\line(1,0){2.0}}
\end{picture}\hspace*{0.3mm}}}

\newcommand{\nv}[0]{
\unitlength=1mm
\linethickness{0.5pt}
\protect{
\begin{picture}(4.40,4.00)
\put(1.2,-0.4){\line(0,1){3.1}}
\put(2.1,-0.4){\line(0,1){3.1}}
\put(2.1,1.1){\line(1,0){2.0}}
\protect{
\put(0.3,-0.7){\line(1,1){3.6}}}
\end{picture}\hspace*{0.3mm}
}
}

\newcommand{\nvv}[0]{
\unitlength=1mm
\linethickness{0.5pt}
\protect{
\begin{picture}(4.40,4.00)
\put(2.1,-0.4){\line(0,1){3.1}}
\put(2.1,1.2){\line(1,0){2.0}}
\protect{
\put(0.6,-0.5){\line(1,1){3.6}}}
\end{picture}\hspace*{0.3mm}
}
}

\newcommand{\dd}[0]{
\rule{1.5mm}{1.5mm}}

\newcommand{\llll}{{\cal LT \hspace*{-0.05cm}L}}

\newcommand{\nn}{{\bf N}}

\newcommand{\pp}{{{\bf N^{-1}}}}

\newcommand{\uu}{{{\bf U}}}

\newcommand{\suu}{{{\bf S}}}

\newcommand{\bb}{{{\bf B}}}

\newcommand{\nnn}{{\cal N}}
\newcommand{\sss}{{{\bf S}}}
\newcommand{\zzz}{{\cal LT \hspace*{-0.05cm}L}_K(Z)}
\newcommand{\zz}{{{\cal Z}_C}}

\date{}

\bigskip

\begin{abstract}
We consider modeling the conception of knowledge in terms of temporal logic.
The study of knowledge logical operations is originated around  1962 by representation of knowledge and belief using modalities. Nowadays, it is very good established area.
 However, we would like to look to it from a bit another point of view, our paper 
  models  knowledge in terms of linear temporal logic with {\em past}.  We consider various versions of logical knowledge operations
which may be defined in this framework.
 Technically, semantics, language and temporal knowledge logics based on our approach are constructed.
 Deciding algorithms are suggested, unification in terms of this approach is commented.
 This paper does not offer strong new technical outputs, instead we suggest
 new approach to conception of knowledge (in terms of time). 
 \end{abstract}

\bigskip


\hspace*{1.8cm}  unification, computation of unifiers, projective formulas,

\hspace*{1.8cm}  deciding algorithms

\bigskip

\section{Introduction}



The conception of knowledge is in a focus of Logic in Computer Science.
 E.g., as a general field, knowledge-representation is a part of artificial intelligence
which is devoted to designing computer representations for capture information about the world that can be used to solve complex problems.
The approach to model knowledge in  terms of symbolic logic, probably, may be dated to the end of 1950.

At 1962 Hintikka \cite{jh4} wrote the book: {\em Knowledge and Belief}, the first book-length work to suggest using modalities to capture the semantics of knowledge.
  This book laid much of the groundwork for the subject, but a great deal of research has taken place since that time. Nowadays the field of knowledge representation and reasoning in logical terms is very wide and active area, it includes,
in particular, modeling of knowledge and common knowledge by tools of agents multi-modal logic.
Frequently  different  variations of
modal and multi-modal logics were used for formalizing agent's reasoning.
Such logics
were, in particular, suggested in Balbiani et al \cite{pb1},
Vakarelov \cite{dv},
Fagin et al
\cite{fag1}, Rybakov \cite{ry2003,vr179}. The book Fagin et al
\cite{fag1} contains summarized to that time systematic approach to study
the notion of common knowledge.
Some contemporary study of knowledge and believes, in particular,  in terms of single-modal logic is contained at Halpern et al \cite{hal1}.

In our paper we would like to discuss a bit another approach to knowledge based at temporal logic (rather than, as earlier,  at epistemic(modal logic), cf. Atremov et al \cite{art1,art2}, Halpern \cite{hal1}).
We will consider knowledge via tools of linear temporal logic LTL, more precisely  via its dual analog with SINCE operation.
This looks (as we hope) very natural and brings various abilities do define knowledge operations and to effectively use it in applications.

The choice of LTL for our approach is, in particular, justified by a big role which LTL plays in CS; yet more important is that usage of LTL forms a core point of our approach, 
as we need operation SINCE to model knowledge.
Historically, temporal logic has been (and is) very active area in mathematical logic and information sciences, CS, etc.
(cf.,  eg,   Gabbay and Hodkinson\cite{ghr,gbho,gbho2}).
The linear temporal logic $ \llll $ (with Until and Next) has important applications in CS (cf.
Manna, Pnueli \ci{ma1,ma2}, Vardi \ci{va1,va2}), e.g. -- for analyzing protocols of computations, check of consistency, etc.
The solution for admissibility problem for $\llll$ itself was found in Rybakov \ci{ry08a}, the basis for admissible rules of $\llll$ was obtained in Babenyshev and Rybakov \ci{babry}
(earlier the case of $\llll$ with no Until was solved in \ci{vr11}; the case of linear temporal logic with future and past easy follows because
we may model in this logic the universal modality (cf. Rybakov \cite{vr11h})).
The solution for unifiability problem in $\llll$ was found
in Rybakov \cite{r12a}. In current our paper we also would like to consider
this problem within accepted framework. 
The unification problem was originated in CS and it
consists of  decision/recognision
if two given
terms may be
transformed into semantically equal ones (via a substitution of other terms in place of variable-letters).

In early stage, unification started as the problem: whether  two given terms may be turned to syntactically equal, by replacing their variables by terms; if yes
they were said to be unifiable. This problem was independently
introduced in automated deduction by Robinson \ci{ro} and in term
rewriting by Knuth et al \ci{kn1}.
Then it was suggested that, instead of making terms syntactically equal,
it is relevant to consider the semantic equivalence: when  all possible values the unified terms would be the same. Since then, all instruments of mathematical logic have been involved in the research concerning this task
(cf. Baader and Snyder \ci{ba4},  Baader and Ghilardi - \ci{bagh}, Baader et al \ci{bamo,ba3,ba4})

Historically, the
unification in intuitionistic logic and propositional modal logics over K4
was intensively studied by
S.~Ghilardi~\cite{r10,g1,ghi1,ghi2,ghi3} (via application ideas from
projective algebras and technique based on projective
formulas).
  In these
works, the problem of construction finite complete sets of unifiers (in logics under
consideration) was solved and computational algorithms were suggested.
This approach gave very useful contributions the problem of admissibility  for inference rules
cf. \ci{ie1,ie2,jer2,jer3,jer4}. The generalized unification problem (for formulas with coefficients) in the intuitionistic logic and modal logic $S4$
was solved in Rybakov \ci{r11a,r11b}.

In works of Ghilardi
a technique of projective formulas was effectively used.
Later this technique was applied in
 Dzik and Wojtylak \cite{dz}, where it was shown that any formula unifiable in the linear modal logic $S4.3$ is projective, which gives a hint that similar might be in a good fragment of $\llll$ (at \cite{dz}, it was fairly observed that ideas similar to projectivity for linear modal and intuitionistic logics
were suggested already in A. Wro´nski \cite{wro1,wro2}). Next, at Rybakov \cite{ry2004} it was shown that
any formula unifiable in the linear temporal logic $\llll_U$ (with only UNTIL) is also projective, and  algorithms to
built unifiers were found.

In this paper we
study the conception of knowledge in terms of temporal logic: how the logical knowledge operations may be
defined (we suggest several plausible versions). 
Technically, semantics, language and temporal knowledge logics were suggested.
Deciding algorithms for them are described (it is just easy applications of known techniques),
and also we consider the unification problem in such logics. From technical standpoint, we just use known results and apply them to suggested logics, so
there are no strong technical new outputs in this paper.
But we see our approach is interesting and important conceptually, because we suggest new natural point of view to knowledge logical operations themselves in terms of temporal logic. We set several
 interesting open problems.

\section{Preliminary Definitions and Notation}

We would like to discuss ways of modeling knowledge (logical knowledge operation) within temporal logic.
Actually, as we noted in the introduction, the approach to representation of knowledge via modality and
other operations of epistemic logic was in study  since about  1962. However
we would like to consider an interpretation of logical knowledge operations via parameterized temporal operations. We will need a technique from the linear temporal logic and modal logics. Therefore we start from a recall of basic definitions and notation.

We will primarily work with a dual of the linear temporal logic $\llll$.
The
language of the Linear Temporal Logic ($ \llll $ in the sequel)
  extends the language of Boolean logic by operations $\nn$ (next)
  and $\uu$ (until).
The formulas of $\llll$ are built up from a set $Prop$ of atomic
 propositions (synonymously - propositional letters)
  and are closed under applications of Boolean
 operations, the unary operation $\nn$ (next) and the binary
 operation $\uu$ (until).
The formula $\nn\ppp$ has meaning: $\ppp$ holds in the next time
point (state); $\ppp \uu \psi$ means: $\ppp$ holds until
$\psi$ will be true.
Standard semantics for $\llll$ consists of {\em infinite
  transition systems (runs, computations)}, formally  they are
  linear Kripke structures based on natural numbers.

The infinite linear Kripke structure is a
 quadruple
 $\cm:=\langle \nnn, \leq, \mathrm{Next}, V
 \rangle$, where $\nnn$ is the set of all natural
 numbers;   $\leq$ is the
  standard order on $\nnn$, $\mathrm{Next}$ is the binary relation,
  where $a \ \mathrm{Next} \ b$ means $b$ is the number
 next to $a$.  And $V$ is a valuation of a subset $S$ of
 $Prop$.

 That is, $V$
 assigns truth values to elements of $S$. So, for any
 $p\in S$, $V(p)\subseteq \nnn$, $V(p)$ is the
 set of all $n$ from $\nnn$
 where $p$ is true (w.r.t. $V$).
The elements of $\nnn$ are {\em states} (worlds), $\leq$ is the {\em
transition} relation (which is linear in our case), and $V$ can be
interpreted as {\em labeling} of the states with atomic
propositions. The triple $\langle \nnn, \leq, \mathrm{Next}
 \rangle$ is a Kripke frame which we will denote for short by
 $\nnn$.



 \smallskip

For any Kripke structure
 $\cm$,
 the truth values can
be extended from propositions of $S$ to arbitrary formulas
constructed from these propositions as follows:

\smallskip

$ \forall p\in Prop \ (\cm,a)\vv_V p \ \lri  a\in \nnn \wedge \ a\in V(p);$

$ (\cm,a)\vv_V (\ppp\wedge \psi) \ \lri $
$(\cm,a)\vv_V \ppp \wedge
(\cm,a)\vv_V \psi;$ 

$ (\cm,a)\vv_V \neg \ppp \ \lri not [(\cm,a)\vv_V \ppp]  ;$

$ (\cm,a)\vv_V \nn \ppp \ \lri \forall b
[(a \ \mathrm{Next} \ b) \ri (\cm,b)\vv_V \ppp]  ;$

$ (\cm,a) \vv_V (\ppp \uu \psi) \ \lri \exists b [
(a\leq b)\wedge ((\cm,b)\vv_V \psi) \wedge $

$
 \forall c [(a\leq c < b) \ri (\cm,c)\vv_V \ppp ]].
$

\smallskip



 { For a Kripke structure $\cm:=\langle \mathcal{N}, \leq,
\mathrm{Next},V
 \rangle$ and a formula $\ppp$ with letters from the domain of $V$,
  we say
$\ppp$ is valid in $\cm$ (denotation -- $\cm\vv \ppp$) if, \ for any
$b$ of $\cm$ ($b\in \mathcal{N}$),
 the formula $\ppp$ is true at $b$ (denotation: $(\cm,b)\vv_V
\ppp)$.}

The linear temporal logic $\llll$ is the set of all formulas
  which are valid in all infinite temporal linear Kripke structures
  $\cm$ based on $\nnn$ with standard $\leq$ and $\mathrm{Next}$.
  The logic $\llll_U$ is the subset of $\llll$ consisting of only formulas
  without $\mathrm{Next} $.

We will basically need a dual of $\llll_U$, the logic with only {\em since} operation.
It may be formulated as follows.
The formulas are constructed as earlier, but with the binary logical operation $\suu$ instead of $\uu$, and without
$\mathrm{Next}$.
The frame $\nnn^{-}$ is $\langle  N,  \geq  \rangle$,
 and $V$ as before is a valuation of a subset $S$ of
 $Prop$ on the set $N$. So, we  take the language of $\llll$, delete $\mathrm{Next}$ and $\uu$ and replace it with the binary operation $\suu$. The definition of the truth relation for
 $\suu$ is as follows:

\[ (\nnn^{-},a) \vv_V (\ppp \suu \psi) \ \lri \exists b [
(b\geq a)\wedge ((\cm,b)\vv_V \psi) \wedge \]

\[
 \hspace*{0.1cm}\forall c [(a \leq c < b) \ri (\cm,c)\vv_V \ppp ]].
 \]
So, $\suu$ is just the dual of $\uu$ (and note pls that it acts exactly as $\uu$, we simply interpret it to {\em past}).

\begin{definition}
The logic $\llll_S^{-}$ is the set of all formula which are true at $\nnn^{-}$ w.r.t.
all valuations.
\end{definition}

The notations and definitions concerning modal logic $S4.3$ are very well known and therefore we omit it; just briefly recall that unary modal operations $\Diamond$ and $\Box$ only are added to the language of Boolean logic.

All linear temporal and modal logics mentioned above are decidable (by any given formula we may compute if this formula belongs to this logic, if it is a theorem); there are many techniques to
construct deciding algorithms. For example, we enumerate below several ones.
First, recall that
a formula $\ppp$ is {\em satisfiable} if there is a structure $\cm$ where
$\ppp$ is true at some world. And a formula $\ppp$ is a theorem iff  $\neg \ppp$
is not satisfiable.
It is well known that
 $\llll$ is decidable and decidable  w.r.t. satisfiability:
for any satisfiable formula we may effectively construct a finite  model (of bounded size)  for  $\ppp$
(it is a standard result of model checking in terms of $\llll$, cf.  \cite{va1,va2}, or, eg, it immediately follows from admissibility technique for $\llll$
suggested in \cite{ry08a}, Lemma 15).
For interested reader, we may remind that such a model is the one with initial part to be a finite interval
of natural numbers (w.r.t. Next)  and with the final part to be a finite cluster - circle - (with a fixed route by Next).
For the logic $\llll_S^{-}$ all said above is true as well, only technique works in the opposite direction - to the past. Now all preliminary information is given, and we are ready to go to logical knowledge operations.

\section{How to Define Logical Knowledge Operations}

It is easy to accept that
 the knowledge is not absolute and depends on opinions of individuals (agents) who
accept a statement as safely true or not, and, yet, on what we actually consider as true knowledge.
We, first,  would like to look at it via temporal perspective.
Some evident trivial observations are that
\bigskip

{\em
(i) Human beings remember (at least some) past, but

\medskip

(ii) they do not know future at all (rather could surmise what will happen in

immediate proximity time step);

\medskip

(iii) individual memory tells to us  that the time in past was linear

(though it might be only our perception).

\medskip

}

Therefore it looks meaningful to look for the interpretation in
 PAST linear temporal logic -  $\llll_S^{-}$.
 We would like to suggest several approaches to define the operation of knowledge:
here we will use the unary logical operations $K_i$ with meaning - it is a knowledge operation.

\medskip

{\bf (i) approach: when knowledge  holds  stable:}

\[ (\nnn^{-},a) \vv_V K_1 \ppp \ \lri \exists b [
(a\geq b)\wedge ((\nnn^{-},b)\vv_V \ppp) \wedge  \]

\[ \forall c [(a\geq c > b) \ri (\nnn^{-},c)\vv_V \ppp ].
\]

 That is

 \[  (\nnn^{-},a) \vv_V K_1 \ppp \ \lri   (\nnn^{-},a) \vv_V  \ppp S \ppp \]

 This is an unusual but rather plausible interpretation.
 In time being, $\ppp$ is a knowledge if one day in past it happened to be true and was true since then until now.

\medskip

{\bf (ii) approach: knowledge  if always  was true}

\[  (\nnn^{-},a) \vv_V K_2 \ppp \ \lri   (\nnn^{-},a) \vv_V \neg ( \top S \neg \ppp) . \]

This is close ro standard interpretation in epistemic logic offered quit a while ago:
we consider a fact to be knowledge if it held always (but  (in our approach) in past).

\medskip

{\bf (iii) approach: via parameterized knowledge}

\[ (\nnn^{-},a) \vv_V K_{\psi} \ppp \ \lri   (\nnn^{-},a) \vv_V  \ppp S \psi \]

This means $\ppp$ has a stable value  true; since some event happened in past, which is modeled now by $\psi$ to be true at a state.
Thus, as soon as $\psi$ happened to be true, $\ppp$ always held to be true. Clearly this approach generalizes first two suggested above  and yet it is more flexible.
From technical standpoint
we just use standard SINCE operation of the linear temporal logic LTL diverted to past.
  But the approach is looking very attractive, since it gives new view angle on the problem and yet uses some respectful and well established technique.

\medskip

{\bf (iv) approach: via agents knowledge as voted truth for the valuation}

\smallskip

This is very well established area, cf. the book Fagin et al
\cite{fag1}
and more contemporary publications
e.g. -  Rybakov \cite{ry2003,vr179}. Though here we would like to look at it from an another standpoint. 
Earlier knowledge operations (agents knowledge) were just unary logical operations
$K_i$ interpreted as $S5$-modalities,
 and knowledge operations were introduced via
the vote of agents, etc.
We would like to suggest here somewhat very simple but anyway rather fundamental, and it seems new.

Here, in order to implement multi-agent's framework we assume that all agents have theirs own valuations at the frame $N^{-}$.
So, we have $n$ much agents, and $n$-much valuations $V_i$, and as earlier the truth values w.r.t. $V_i$ of any propositional letter $p_j$ at any world $a\in N$.
For applications viewpoint, $V_i$
correspond to agents information about truth of $p_j$ (they may be different).
So, $V_i$ is just individual {\em information} .

How the information can be turned to {\em local} knowledge?
One way is the voted value of truth: we consider a new valuation $V$, w.r.t. which  $p_i$ is true at $a$ if majority, biggest part of agents,
believes that $p_i$ is true at $a$. Then we achieve a model with a single (standard) valuation $V$. Then we can apply any of proposed upper approaches
to introduce, so to say, logical operations of global knowledge $K$ as it has been shown above.

\medskip

{\bf (v) approach: via agents knowledge as conflict resolution

\ \ \ at evaluating point}

\smallskip

Here we suggest a way starting similar as in the case (iv) above until introduction of
different valuations $V_i$ of agent's opinion.
But then we suggest

 \[  (\nnn^{-},a) \vv_V K \ppp \ \lri   \forall i [(\nnn^{-}, a)  \vv_{V_i}  \ppp S \ppp ]. \]

 In this case, if we will allow nested knowledge operations together with several valuations $V_i$ for agent's information and yet derivative valuation $V$ for all cases when we
 evaluate $K\ppp$ (regardless for which agent (i.e. $V_i$)),
 no decision procedure is known. We think that to study it is an interesting open question.

  Yet one open interesting question is to extend the suggested approach to linear logic based at all integer numbers $Z$  (which means we have infinite past and infinite future). Then the knowledge will be interpret by a stable truth on a reasonable interval of time in past and future, so we will need to use both
operations $\suu$ and $\uu$. Next, it would be good to study such approach to the case of continuous time (in past, or both - past and future).
One more open, and it seems not easy question, is how to extend results of this paper (including unification from next section)
to the linear temporal logic with the analog of Next operation directed to past.  This completes our section with
suggestions of various interpretations for logical knowledge operations.

\section{Appendix:  Unifiability Problem}

This part is a bit apart from general line of this paper, because we
would like to comment unification problem in this framework. Unification does not much directly to just conception of knowledge immediately (though potentially is interesting), but the author likes unification problem nowadays.
Fortunately all necessary technical results were obtained recently, and we need only to point how to apply them. We start from a recall of definitions concerning unifiability.

\begin{definition}
A formula $\ppp$ is unifiable in a logic $L$ if there is a substitution
$\varepsilon$ (which is called a unifier for $\ppp$) such that $
\varepsilon(\ppp) \in L$.
A unifier $\varepsilon$ (for a formula $\ppp$ in a logic $L$) is more
general than an another unifier $\varepsilon_1$ iff there is a substitution
$\delta$ such that for any letter $x$, $[\varepsilon_1(x) \equiv
\delta(\varepsilon(x))] \in L$.
\end{definition}



To check just the unifiability of a formula in $L$ (if $L$ is decidable)
is (theoretically, not computationally) an easy task: it is sufficient
to use only ground substitutions: mappings of variable-letters in the set
$\{\bot, \top\}$. But the problem - how to find all unifiers -
all solving substitutions - is not easy at all.
Below we just recall, transform and implement results from \cite{ry2004}

\begin{definition}
 A set of unifiers $CU$ for a given formula $\ppp$ in a logic $L$ is a complete set of
unifiers, if the following holds. For any unifier
$\sigma$ for $\ppp$ in $L$, there is a unifier $\sigma_1$ from $CU$, where $\sigma_1$ is more general than
$\sigma$.
\end{definition}

Since the logic $\llll_{S}^{-}$ itself, has definable $\Box^- $ and $\Diamond^- $, - $ \Diamond^- x :=  \top \suu x $, $\Box^- = \neg \Diamond^- \neg$,  we may formulate projectivity as follows:

\begin{definition} A formula $\ppp$ is said to be projective in
$\llll_{S}^{-}$ if the following holds. There is a substitution $\sigma$ (which is called
projective substitution) such that $\Box^- \ppp \ii [ x_i \equiv \sigma (x_i)] \in \llll_{S}^{-}$ for any letter $x_i$ from $\ppp$.
\end{definition}

\begin{lemma} (remake from \cite{ry2004}) \label{l1} If a substitution $\sigma_p$ is projective for a formula $\ppp$  $\llll_{S}^{-}$, then the set $\{\sigma_p\}$ is a complete set of unifiers for $\ppp$ (i.e. $\sigma_p $ is most general unifier for $\ppp$).
\end{lemma}

\pr. Indeed, let $\sigma$ be a unifier for $\ppp$ in $\llll_{S}^{-}$.
Since we assume $\sigma_p$ is projective for $\ppp$ in $\llll_{S}^{-}$, we have
$\Box^- \ppp \ii [x_i \equiv \sigma_p(x_i)] \in \llll_{S}^{-}$ for any letter $x_i$ from $\ppp$. Acting by $\sigma$ on the formula above we get
$\sigma(\Box^- \ppp) \ii [\sigma(x_i) \equiv \sigma (\sigma_p (x_i))] \in \llll_{S}^{-} $,
that is $\sigma(x_i) \equiv \sigma (\sigma_p (x_i)) \in \llll_{S}^{-} $. $\Box$

\medskip

\begin{theorem} \label{mntA}
Any formula $\ppp$ unifiable in $\llll_S^{-}$  is projective and we may compute a projective unifier for $\ppp$.
\end{theorem}

We can prove this theorem exactly the same way as for the logic $\llll_U$ in Rybakov
\cite{ry2004}. It is just remake of that proof by replacing {\em future by past.}
$\Box$

\medskip

Using this theorem we also immediately obtain a solution for admissibility problem. The point is,
as soon as we
 possess a projective unifier, it is sufficient to use only it to check if a rule is admissible. So, we have more as just decidability algorithm for the logic itself.
This approach covers all suggested interpretations (i) - (iv) for logical knowledge operation suggested above. The case (v) is open.

\section {Conclusion}

Our paper considers
 the conception of {\em knowledge } in terms of linear temporal logic.
We suggest several ways to define logical knowledge operations in chosen framework.
Semantics, language and temporal knowledge logics based at our approach are suggested.
We also consider the unification problem and
show how to solve it.
The paper concentrates primarily on the conception of knowledge itself,
suggests several new ways to study the problem, a new view angle on definitions.
We set also several interesting open problems.


\end{document}